# QuantumCumulants.jl: A Julia framework for generalized mean-field equations in open quantum systems

David Plankensteiner, Christoph Hotter, and Helmut Ritsch

Institut für Theoretische Physik, Universität Innsbruck, Technikerstr. 21a, A-6020 Innsbruck, Austria

A full quantum mechanical treatment of open quantum systems via a Master equation is often limited by the size of the underlying Hilbert space. As an alternative, the dynamics can also be formulated in terms of systems of coupled differential equations for operators in the Heisenberg picture. This typically leads to an infinite hierarchy of equations for products of operators. A well-established approach to truncate this infinite set at the level of expectation values is to neglect quantum correlations of high order. This is systematically realized with a so-called cumulant expansion, which decomposes expectation values of operator products into products of a given lower order, leading to a closed set of equations. Here we present an open-source framework that fully automates this approach: first, the equations of motion of operators up to a desired order are derived symbolically using predefined canonical commutation relations. Next, the resulting equations for the expectation values are expanded employing the cumulant expansion approach, where moments up to a chosen order specified by the user are included. Finally, a numerical solution can be directly obtained from the symbolic equations. After reviewing the theory we present the framework and showcase its usefulness in a few example problems.

https://github.com/qojulia/QuantumCumulants.jl

## 1 Introduction

Numerical simulation of the time evolution of quantum systems has become a key method in the fields of quantum optics and quantum information as analytic approaches and solutions are only rarely available. In particular, for open systems one has to go beyond the determination of eigenenergies and eigenstates of a given Hamiltonian and look for the stationary solution of the corresponding system density matrix. To study processes like quantum annealing even solutions for time-dependent Hamiltonians are required. As there are only a finite number of building blocks for quantum optics Hamiltonians, the desire to reduce the repeated effort for implementing quite similar problem Hamiltonians and master equations has led to the development of several generic numerical frameworks and developer tools in the past decades. These simultaneously target fast coding, high memory efficiency and short execution times.

Among the most well-known frameworks are the early Quantum Optics MATLAB toolbox implemented by Sze Tan already over two decades ago [1] and its first open source successor, the Quantum Toolbox in Python (QuTiP) [2, 3]. Further efforts to optimize code efficiency and memory use lead to C++ based software such as C++QED [4]. Similar projects exist for quantum information applications and highly correlated quantum spin systems [5, 6, 7, 8].

More recently, based on the highly efficient Julia programming language [9], the framework QuantumOptics.jl [10] was developed with the goal to implement a large class of quantum optics problems and methods. All these numerical toolboxes aim to streamline and simplify the usage of standard numerical techniques, which are generally based on the matrix representation of quantum mechanical operators in truncated Hilbert spaces. Equations of motion that govern the dynamics of quantum systems, such as the Schrödinger equation for closed systems, stochastic wavefunction simulations or the master equation for open systems then amount to solving finite sets of (stochastic) coupled differential equations.

One practical limitation here is the growing size of the matrices representing quantum mechanical operators in larger Hilbert spaces. A generic example is the exponential growth of the size of the Hilbert space describing $N$ two-level atoms (qubits) whose state is described by a vector of a size scaling as $2^N$. Correspondingly, operators in this Hilbert space are represented by matrices of the size $2^N \times 2^N$. Clearly, representing many atoms is hardly feasible in such an approach. In some specific cases this number has been pushed to several tens of spins [11], which is still well below the numbers nowadays available in experimental implementations.

In order to treat configurations of higher numbers of subsystems, one has to apply different techniques to reduce the problem size. One highly successful approach in the study of lattice dynamics is based on matrix product states and the density renormaliza-tion group [12, 13]. Given a state that describes a potentially large composite system one can reduce the degrees of freedom to the relevant ones to represent the state vector more efficiently thereby reducing the dimensionality of the problem. Similar restrictions to the low energy sector of a many-body Hilbert space can lead to sufficient accuracy in the low temperature or low excitation regime.

An alternative approach follows a similar idea by neglecting quantum correlations of higher order leading to a sufficiently accurate description in regimes exhibiting low correlations. The starting point are the $c$-number differential equations describing the dynamics of the expectation values of a given set of operators. These are generally coupled to higher order products of these operators leading to an infinite number of equations. In order to truncate the set of equations one can systematically approximate the higher-order products. To this end the concept of cumulants, originally conceived for the treatment of stochastic variables, has been generalized to operators [14]. The joint cumulant of a set of operators is a measure of the correlations of those operators, in that it vanishes if one (or a subset) of the operators is statistically independent of the others. Neglecting quantum correlations above a certain order is then equivalent to neglecting the joint cumulant of a set of operators.

Such order-based reductions allow for the construction of a closed set of ordinary differential equations for the averages of operators of interest. The equations of motion for the corresponding averages are obtained from the operators which appear as noncommutative variables, that obey prescribed fundamental commutation relations. These commutation relations can be used in the symbolic computation of the Heisenberg equations (for closed systems) or Quantum Langevin equations (for open systems) in operator form, from which the equations of motion for averages can be obtained. Neglecting the joint cumulants above a certain order generally leads to a complexity that scales polynomially, where the leading order is determined by the point up to which correlations are kept. In this sense a second-order treatment of $N$ identical subsystems requires numerically solving $\mathcal{O}(N^2)$ equations.

As already discovered in the early days of quantum optics, the generalized cumulant expansion approach works well in open-system problems such as lasing [15, 16]. In the recent renewed interest in so-called superradiant lasers, where large ensembles of particles interact via a single cavity mode, the method was extensively used with notable success [17, 18, 19, 20, 21, 22, 23, 24, 25, 26]. Furthermore, not only cavity QED systems have been calculated, but also ensembles of interacting free-space atoms, pointing out the importance of higher orders in specific cases [27, 28, 29].

The disadvantage of this method is that it requires the analytical computation of a large number of equations even at fairly low orders of expansions and thus lacks the conceptual simplicity of encoding the numerical solution of master equations via the toolboxes mentioned in the beginning. Higher-order treatments can quickly result in a set of equations that is too large to be manageable by hand in a reasonable amount of time. Furthermore, for each modification one wants to implement with this approach it is required to rederive the equations of motion for the operators before performing the cumulant expansion up to the desired order.

In this paper, we present a framework which aims to address these issues. We developed QuantumCumulants.jl in order to automate the derivation of equations, the cumulant expansion to an arbitrary order, as well as the final step of numerically implementing the resulting set of equations. It is written in the Julia language [9], combining a certain ease of use with high performance. Before presenting the framework itself, we will provide a review of the theory behind the cumulant expansion approach [14]. Next, we will highlight some details of the implementation. After that, we showcase the capability of the framework at the hand of some examples. Finally, we provide an overview of the framework's limitations and an outlook on possible future changes.

## 2 Theoretical background

In this section we will set the ground for the programmatic framework by providing a detailed review of the underlying theory.

### 2.1 A brief example

For didactic purposes, we will specify a model of a well-known example for which the cumulant expansion approach is known to work well. We will return to this example at each step of the approach for clarity.

Consider the simplest quantum model of a laser: a single two-level atom, acting as gain medium, is placed inside an optical cavity, cf. Fig. 1(a). The coherent dynamics of this system is described by the Jaynes-Cummings Hamiltonian,

$$H_{\mathrm{JC}} = \hbar\Delta a^\dagger a + \hbar g \left(a^\dagger \sigma^{ge} + a\sigma^{eg}\right). \quad (1)$$

Here, $\Delta = \omega_{\mathrm{c}} - \omega_{\mathrm{a}}$ is the detuning between the cavity resonance frequency $\omega_{\mathrm{c}}$ and the atomic transition frequency $\omega_{\mathrm{a}}$. The coupling between the cavity and the atom is governed by the frequency $g$. The operators $\sigma^{ij} = |i\rangle\langle j|$ describe atomic transitions between the ground state $|g\rangle$ and its excited state $|e\rangle$, whereas the field dynamics are characterized by the photonic annihilation operator $a$ and the creation operator $a^\dagger$.

The atom can spontaneously emit a photon at a rate $\gamma$, described by the damping operator $\sigma^{ge}$. At the same time, the cavity loses photons at a rate $\kappa$, which is included with the damping operator $a$. Finally, gain to the system is provided by incoherently driving the atom from the ground to the excited state at rate $\nu$ with the "damping" operator $\sigma^{eg}$ [see Fig. 1(a)].

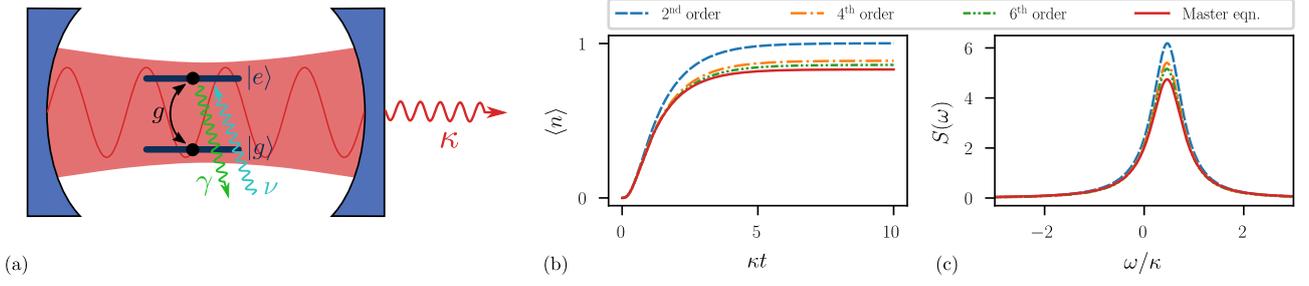

Figure 1: *The single-atom laser.* (a) Schematic illustration of an optical resonator containing a single atom that acts as a gain medium for the laser. (b,c) Comparison of the dynamics when treating the single-atom laser model in a second-, fourth- and sixth-order approximation, as well as in a full quantum model (master equation). In (b), we show the average photon number, and (c) shows the power density spectrum of the laser. The parameters where $\Delta = \kappa/2$, $g = 1.5\kappa$, $\gamma = 1.25\kappa$, and $\nu = 4\kappa$.

Solving the Master equation of this system will yield the time dynamics of the density matrix, which contains the entire information about the system. However, if we only want to compute some specific expectation values in the end, we may not need all this information. Rather, the task is then to find equations of motion for the averages that allow us to solve for them directly. As we will see in the following, this can be achieved by approximating the full dynamics, which also reduces the numerical complexity of obtaining the average values.

## 2.2 Equations of motion in the Heisenberg picture

The starting point of any system that is to be simulated is its definition. Essentially, this means that the Hamiltonian and corresponding decay channels (if any) are specified. The time evolution of a system operator $\mathcal{O}$ is then given by

$$\dot{\mathcal{O}} = \frac{i}{\hbar}[H, \mathcal{O}] + \sum_n [c_n^\dagger, \mathcal{O}] \left( \sqrt{\gamma_n} \xi_n(t) + \frac{\gamma_n}{2} c_n \right) - \left( \sqrt{\gamma_n} \xi_n^\dagger(t) + \frac{\gamma_n}{2} c_n^\dagger \right) [c_n, \mathcal{O}]. \quad (2)$$

In the above, $H$ is the Hamiltonian describing the unitary time evolution of the system. Additionally, the system features distinct decay channels with rates $\gamma_n$ and corresponding collapse operators $c_n$. The operators $\xi_n$ describe quantum noise. Eq. (2) is known as the Quantum Langevin equation [30].

In the following, we will assume that any noise that occurs in the systems is white noise. Then, the operators $\xi_n$ in Eq. (2) do not contribute to any averages we compute. Hence, if we are only interested in averages

we may drop the noise terms from Eq. (2) and compute

$$\dot{\mathcal{O}} = \frac{i}{\hbar}[H, \mathcal{O}] + \sum_n \frac{\gamma_n}{2} \left( 2 c_n^\dagger \mathcal{O} c_n - c_n^\dagger c_n \mathcal{O} - \mathcal{O} c_n^\dagger c_n \right). \quad (3)$$

For closed systems, i.e. $\gamma_n = 0$, we recover the Heisenberg equation. In the following, we will develop a theory that allows us to obtain a closed set of $c$-number equations from Eq. (3) by truncating the "order" of averages (or moments) of operator products.

To illustrate, take the example of a single-atom laser. Say we would like to compute the time evolution of the average of the field operator $a$. Using Eq. (2) in order to derive the corresponding operator equation of motion, we see that $a$ depends on the atomic operators as well. Hence, we derive the equations for all occurring operators to obtain a complete set of equations, and find

$$\dot{a} = -\left(i\Delta + \frac{\kappa}{2}\right) a - ig\sigma^{ge}, \quad (4a)$$

$$\dot{\sigma}^{ge} = -\frac{\gamma + \nu}{2} \sigma^{ge} + iga\sigma^{ee}, \quad (4b)$$

$$\dot{\sigma}^{ee} = -\gamma \sigma^{ee} + \nu(1 - \sigma^{ee}) + ig\left(a^\dagger \sigma^{ge} - a\sigma^{eg}\right). \quad (4c)$$

The above set of equations consists of operator equations. Solving these directly is equivalent of the numerical complexity of solving a Master equation. In addition, Eq. (4) is incomplete: in order to obtain correct results, we need to include the quantum noise; i.e., the equations are stochastic operator differential equations, making them numerically even more expensive.

Averaging over the above equations, however, makes them more tractable. The averages of operators are simple $c$-numbers and we hope to obtain a set of equations that is simple to solve in the end. For the averages

of Eq. (4), we find

$$\langle \dot{a} \rangle = -\left(i\Delta + \frac{\kappa}{2}\right)\langle a \rangle - ig\langle \sigma^{ge} \rangle, \quad (5a)$$

$$\langle \dot{\sigma}^{ge} \rangle = -\frac{\gamma + \nu}{2}\langle \sigma^{ge} \rangle + ig\langle a\sigma^{ee} \rangle, \quad (5b)$$

$$\langle \dot{\sigma}^{ee} \rangle = -\gamma\langle \sigma^{ee} \rangle + \nu(1 - \langle \sigma^{ee} \rangle) + ig\left(\langle a^\dagger \sigma^{ge} \rangle - \langle a\sigma^{eg} \rangle\right). \quad (5c)$$

Now, we are confronted with another problem: averages of operator products, such as $\langle a\sigma^{ee} \rangle$ occur in Eq. (5), and since in general $\langle a\sigma^{ee} \rangle \neq \langle a \rangle \langle \sigma^{ee} \rangle$ the set of $c$-number equations is incomplete. Deriving equations of motion for those averages does not solve the problem either, since they will couple to averages of ever longer operator products. We thus converted the operator equations to $c$-number equations. However, this does not lead to a closed set of equations and involves contributions of any order. Hence, in order to find a full solution one would have to derive and solve infinitely many equations.

We therefore have to resort to some kind of approximation which allows for a cutoff at higher orders. The systematic approach to do this the so-called cumulant expansion [14] which we will review in the next section.

## 2.3 Cumulant expansion

At this point of the theoretical introduction, we will introduce the most essential part of the approach used in our simulation package QuantumCumulants.jl. To this end, we provide a short review of the generalized cumulant expansion method, which was initially introduced by R. Kubo [14]. The basic idea is to truncate $c$-number equations such as Eqs. (5) by expanding averages of higher-order operator products in terms of products of lower order expectation values. To clarify, we call the number of operator constituents in a product the "order" of the product; e.g., $\langle a\sigma^{ee} \rangle$ is of order 2.

The joint cumulant (which we denote by $\langle \cdot \rangle_c$) of a product of order $n$ of the operators $\{X_1, ..., X_n\}$ is given by [14]

$$\langle X_1 X_2 ... X_n \rangle_c = \sum_{p \in P(\mathcal{I})} (|p|-1)!(-1)^{|p|-1} \prod_{B \in p} \left\langle \prod_{i \in B} X_i \right\rangle. \quad (6)$$

In the above, $\mathcal{I} = \{1, 2, ..., n\}$, $P(\mathcal{I})$ is the set of all partitions of $\mathcal{I}$, $|p|$ denotes the length of the partition $p$, and $B$ runs over the blocks of each partition. As a short example, consider $n = 3$, where we have

$$\langle X_1 X_2 X_3 \rangle_c = \langle X_1 X_2 X_3 \rangle \\ - \langle X_1 X_2 \rangle \langle X_3 \rangle - \langle X_1 X_3 \rangle \langle X_2 \rangle \quad (7) \\ - \langle X_1 \rangle \langle X_2 X_3 \rangle + 2\langle X_1 \rangle \langle X_2 \rangle \langle X_3 \rangle.$$

Let us stress here that the cumulant of order $n$ is given by averages of order $n$ or lower. Furthermore, the average of order $n$ occurs precisely once on the right-hand-side of Eq. (6).

The joint cumulant can be thought of as a general measure of correlation of operators. The key assumption behind the cumulant expansion approach follows from Theorem I in Ref. [14]. This theorem states that the joint cumulant of a set of operators is zero if any one (or any subset) of them is statistically independent of the others. The assumption we are making is to essentially invert this statement: instead of computing the joint cumulant of a given order to see if it is zero, we assume that it is. Since the average of the same order as the cumulant occurs only once in Eq. (6), we may rearrange the relation to arrive at an expression of an average in terms of only lower-order averages; i.e., if we assume the joint cumulant of order $n$ to vanish, $\langle X_1 X_2 ... X_n \rangle_c = 0$, then the average of the same order is given by

$$\langle X_1 X_2 ... X_n \rangle = \sum_{p \in P(\mathcal{I}) \setminus \mathcal{I}} (|p|-1)!(-1)^{|p|} \prod_{B \in p} \left\langle \prod_{i \in B} X_i \right\rangle, \quad (8)$$

where now $P(\mathcal{I}) \setminus \mathcal{I}$ is the set of all partitions of $\mathcal{I}$ that does not contain $\mathcal{I}$ itself. We therefore approximate the average of order $n$ by an expression that only involves averages of the order $n-1$ and below.

Returning to the example of $n = 3$, we see that when assuming $\langle X_1 X_2 X_3 \rangle_c = 0$ in Eq. (7), we arrive at

$$\langle X_1 X_2 X_3 \rangle = \langle X_1 X_2 \rangle \langle X_3 \rangle + \langle X_1 X_3 \rangle \langle X_2 \rangle \quad (9) \\ + \langle X_1 \rangle \langle X_2 X_3 \rangle - 2\langle X_1 \rangle \langle X_2 \rangle \langle X_3 \rangle.$$

Note that the joint cumulant of order $n = 2$ is just the covariance. Neglecting this is equivalent to neglecting all quantum correlations in the system leading to the assumption that $\langle X_1 X_2 \rangle = \langle X_1 \rangle \langle X_2 \rangle$. This is the lowest-order mean-field approximation and renders the dynamics of the system under consideration classical.

The relation obtained in Eq. (8) can also be applied recursively, if for example one aims to express a fourth-order average as second-order terms only. Applying Eq. (8) once yields an expression consisting only of third-order averages and below. Upon repeated application, we can express the third-order terms as second-order ones, finally leading to an expression that has no terms above the second order.

Let us now return to our example of the single-atom laser. As we have just seen, the simplest thing we can do is a first-order cumulant expansion in which we neglect all correlations. The set of equations from Eq. (5) then becomes

$$\langle \dot{a} \rangle = -\left(i\Delta + \frac{\kappa}{2}\right)\langle a \rangle - ig\langle \sigma^{ge} \rangle, \quad (10a)$$

$$\langle \dot{\sigma}^{ge} \rangle = -\frac{\gamma + \nu}{2}\langle \sigma^{ge} \rangle + ig\langle a \rangle \langle \sigma^{ee} \rangle, \quad (10b)$$

$$\langle \dot{\sigma}^{ee} \rangle = -\gamma\langle \sigma^{ee} \rangle + \nu(1 - \langle \sigma^{ee} \rangle) - 2g\,\mathrm{Im}\left\{\langle a^\dagger \rangle \langle \sigma^{ge} \rangle\right\}. \quad (10c)$$

The above system of equations is simple enough to solve using standard numerical techniques. However, it does not capture any quantum mechanical properties of the lasing setup. Furthermore, we can also see that if we consider the system to initially have zero field ($\langle a \rangle = 0$), and no coherence stored in the atom ($\langle \sigma^{ge} \rangle = 0$) we will never observe any lasing action. This is a signature of the phase invariance of the considered system: in a full quantum treatment, phase-dependent terms such as $\langle a \rangle$ are 0. Only phase-independent operators are nonzero in this example.

We proceed by employing the cumulant expansion approach to obtain a second-order treatment of the single-atom laser in hopes of a more accurate description. Just as before, we derive a set of operator equations, but this time operators such as $a^\dagger a$ are considered. Averaging over the resulting equations, we find

$$\frac{d}{dt} \langle a^\dagger a \rangle = -ig \langle a^\dagger \sigma^{ge} \rangle + ig \langle a \sigma^{eg} \rangle - \kappa \langle a^\dagger a \rangle, \tag{11a}$$

$$\frac{d}{dt} \langle a^\dagger \sigma^{ge} \rangle = \left( i\Delta - \frac{\gamma + \nu + \kappa}{2} \right) \langle a^\dagger \sigma^{ge} \rangle + ig \left( \langle \sigma^{ee} \rangle - \langle a^\dagger a \rangle \right) + 2ig \langle a^\dagger a \sigma^{ee} \rangle, \tag{11b}$$

$$\frac{d}{dt} \langle \sigma^{ee} \rangle = -\gamma \langle \sigma^{ee} \rangle + \nu \left( 1 - \langle \sigma^{ee} \rangle \right) + ig \left( \langle a^\dagger \sigma^{ge} \rangle - \langle a \sigma^{eg} \rangle \right). \tag{11c}$$

The only term that keeps the above set of equations from being closed is $\langle a^\dagger a \sigma^{ee} \rangle$. Assuming that these operators are sufficiently uncorrelated, we employ the cumulant expansion from Eq. (9) to find

$$\begin{aligned}\langle a^\dagger a \sigma^{ee} \rangle &= \langle a^\dagger a \rangle \langle \sigma^{ee} \rangle + \langle a^\dagger \sigma^{ee} \rangle \langle a \rangle \\ &\quad + \langle a^\dagger \rangle \langle a \sigma^{ee} \rangle - 2 \langle a^\dagger \rangle \langle a \rangle \langle \sigma^{ee} \rangle \\ &= \langle a^\dagger a \rangle \langle \sigma^{ee} \rangle. \end{aligned} \tag{12}$$

In the second step we used that all phase-dependent averages vanish. Using the above completes the set of equations from Eqs. (11), and we can once again numerically solve it in a straightforward fashion. We find that it indeed provides a more precise description of the system, nicely approximating the full quantum model, as shown in Fig. 1(b) and Fig. 1(c).

For comparison, we also plot higher-order approximations in Fig. 1(b) and Fig. 1(c). You can see that with increasing order the approximate results converge to the full quantum description. From that we deduce that the approach based on a cumulant expansion is suitable for the description of physical systems. We note, however, that the overall accuracy of the approximations and convergence with higher orders is strongly dependent on the chosen parameter regime. Going into a strong coupling regime (increasing the coupling $g$), for example, would lead to a high degree of quantum correlations between the atom and the cavity mode. Approximations in low order will fail to capture these. A sufficiently accurate approximation, however, might not be feasible since high orders lead to a large number of equations that need to be considered.

The single-atom laser considered here is already well-studied and understood. What if we wanted to adapt the model in order to describe a more realistic lasing setup? This could be done by, for example, adding multiple levels to the atom and considering a more realistic driving scheme. A laser usually features a gain medium consisting of a large number of atoms. What if we wanted to consider many atoms? The cumulant expansion approach is well suited for this as the complexity scales polynomially rather than exponentially with the number of atoms as in a full quantum treatment. However, to actually investigate any changes in the model, we would have to rederive all equations of motion, perform the cumulant expansion, complete the system of equations, and finally implement it for numerical solution. Each step involved in this procedure is both error-prone and tedious.

## 3 The framework

The entire procedure explained above can be automatically performed by QuantumCumulants.jl. To summarize, here is the step-by-step approach with which problems can be treated within the framework:

1. The Hilbert space of the system under consideration is defined. This is necessary since operators acting on different Hilbert spaces commute.

2. Fundamental operators and the corresponding Hamiltonian and dissipative processes need to be defined.

3. Equations of motion for a given set of operators are derived.

4. These equations are averaged and expanded in terms of cumulants up to a specified order.

5. A numerical solution can be obtained directly from the symbolic set of equations.

In this section we will provide a detailed description of the basic concepts QuantumCumulants.jl uses to perform these steps. For more extensive instructions on the usage of the framework, we refer the reader to the documentation that is available online [31]. The source code is also publicly accessible and hosted on GitHub [32].

We note that the noncommutative algebra and the symbolic rewriting using commutation relations is directly implemented in QuantumCumulants.jl. Standard simplifications are performed with the Symbolics.jl [33] library. In the final step, the modeling framework ModelingToolkit.jl [34] is employed to generate fast numerical code that can be used to obtain a numerical solution with the DifferentialEquations.jl [35] package.

## 3.1 Hilbert spaces and operators

The first thing one has to specify when treating a system is the Hilbert space that describes this system. There are currently two Hilbert spaces implemented: one that represents the quantum harmonic oscillator, a Hilbert space of infinite dimension, called `FockSpace`. The other one describes a finite set of discrete energy levels (such as atoms) and is called `NLevelSpace` as it allows for arbitrarily many energy levels. Note that there is no principal limitation that prevents the implementation of other kinds of Hilbert spaces, but these fundamental two cover a large number of problems already (see also Sec. 5.2).

A Hilbert space is a complete vector space. Choosing a basis in such a vector space allows one to represent operators as matrices. This is commonly used in order to solve Master equations. On a more abstract level, however, one can omit the choice of a basis, denoting operators as noncommutative elements of the Hilbert space rather than as matrices. This latter approach is used by QuantumCumulants.jl: operators are defined as noncommutative variables on a specified Hilbert space. Algebraic combinations of operators (such as addition and multiplication) are only possible if the combined operators are defined on the same Hilbert space. To treat composite systems, one needs to consider the product of the respective Hilbert spaces. Consider, for example, the Jaynes-Cummings model. Let $\mathcal{H}_c$ be the Hilbert space of the cavity, and $\mathcal{H}_a$ the Hilbert space of the atom, respectively. The operators in the Jaynes-Cummings model are then defined as elements of the Hilbert space $\mathcal{H}_c \otimes \mathcal{H}_a$. Strictly speaking, we would have to define

$$a \equiv a_0 \otimes \mathbb{1}_a, \quad (13)$$
$$\sigma^{ij} \equiv \mathbb{1}_c \otimes \sigma_0^{ij}, \quad (14)$$

where $\mathbb{1}_i$ is the identity operator on the Hilbert space $\mathcal{H}_i$. Technically, $a_0$ is the photonic annihilation operator, and $a$ is only its extension on the product space. This is necessary since, e.g. the product $a_0 \sigma_0^{eg}$, is not defined, yet the product $a\sigma^{eg}$ is.

This rigorous distinction is usually omitted since the action of $a$ on the atomic Hilbert space is trivial (and vice-versa for $\sigma$). However, we have to take this into account when implementing algebra necessary for operators: while they are noncommutative in general, they commute if they act trivially on disjunct subspaces. Therefore, QuantumCumulants.jl stores the information on which subspace each operator acts nontrivially and uses it to swap operators where it is allowed to do so. In that sense, the operators of Eq. (13) are stored as, e.g. $a \in \mathcal{H}_c \otimes \mathcal{H}_a$ acting nontrivially on 1 (the first subspace) and $\sigma \in \mathcal{H}_c \otimes \mathcal{H}_a$ acting nontrivially on 2.

```
# Load the package
using QuantumCumulants

# Define Hilbert spaces and product space
hc = FockSpace(:cavity)
ha = NLevelSpace(:atom,(:g,:e))
h = hc ⊗ ha

# Define the operators
a = Destroy(h,:a)
σge = Transition(h,:σ,:g,:e)
```

Code sample 1: *Defining the Hilbert space and the fundamental operators of the Jaynes-Cummings model.*

In order to treat any system in the framework, one therefore has to specify the respective Hilbert spaces and then the operators. Note, that QuantumCumulants.jl infers the spaces on which an operator acts nontrivially if the choice is unambiguous. In the case of the Jaynes-Cummings model, the photonic annihilation operator can only be defined on the first Hilbert space, as the other one is not a space representing a quantum harmonic oscillator. If there are multiple Hilbert spaces of the same type in a composite system, however, the space on which an operator acts nontrivially must be explicitly specified on construction (see for example code sample 5).

## 3.2 Application of commutation relations

The basic simplification in the framework uses a few fundamental commutation relations which are immediately applied in any calculation involving operators. On top of that, standard algebraic simplification is performed using the Symbolics.jl [33] framework. In the following, we detail which commutation relations are used. Note that we rewrite terms such that the result adheres to normal ordering. The bosonic annihilation and creation operators fulfil the canonical commutation relation

$$[a, a^\dagger] = 1, \quad (15)$$

which is implemented such that all occurrences of the product $aa^\dagger$ are replaced by $a^\dagger a + 1$. Products of operators describing transitions between discrete energy levels are computed as

$$\sigma^{ij}\sigma^{kl} = \delta_{jk}\sigma^{il}, \quad (16)$$

which either vanishes or results in another transition operator. An additional property of these operators is that the sum over all projectors is conserved and equal to unity. I.e, considering $n$ levels, we have

$$\sum_{i=1}^{n} \sigma^{ii} = 1. \quad (17)$$

This property is used to reduce the number of equations by replacing one of the projectors. By default, the projector of the first level specified is replaced in the framework.

As mentioned above, the commutation relations are applied as soon as operators are combined in a multiplication. This is a deliberate design choice since it is imperative that all possible commutation relations are applied prior to the cumulant expansion. For example, if we were to perform a first-order cumulant expansion on the term $aa^\dagger$, we would have

$$\langle aa^\dagger \rangle = \langle a^\dagger a \rangle + 1 \approx |\langle a \rangle|^2 + 1. \qquad (18)$$

If the commutation relations were not applied at some point, the cumulant expansion would instead yield

$$\langle aa^\dagger \rangle \approx |\langle a \rangle|^2. \qquad (19)$$

Obviously, this is incorrect as it would imply that $\langle aa^\dagger \rangle = \langle a^\dagger a \rangle$.

## 3.3 Averaging and cumulant expansion

On the one hand, computing the average of an operator is straightforward: it is simply converted to a complex number. The same is done for products of operators. For constants involved in the product, as well as for addition, the linearity is used. For example,

$$\langle \lambda_1 ab + \lambda_2 c \rangle = \lambda_1 \langle ab \rangle + \lambda_2 \langle c \rangle, \qquad (20)$$

where $a$, $b$, and $c$ are operators and $\lambda_i \in \mathbb{C}$.

On the other hand, the cumulant expansion is more involved, as has been described in Sec. 2. The framework implements the expansion using a programmatic version of Eq. (8). To this end, standard combinatoric functions from the mathematical Combinatorics.jl library are employed [36].

## 3.4 Additional features

Here, we will provide a brief overview of some convenient features that QuantumCumulants.jl offers.

### 3.4.1 Automatic completion of systems

So far, we have dealt only with the single-atom laser. In second-order, this led to a comparatively small number of equations. It was thus easy to see which equations of motion were necessary in order to arrive at a complete set of equations. However, when dealing with many equations, this might no longer be so simple. Even for small systems, though, we would like to avoid the iterative work of looking for averages that are missing from the equations, and adding them to the set.

The framework offers an automized version of this procedure: using the `complete` function, it will look for any averages that occur on the right-hand-side in a system of equations and check whether the equations of motion are already there. If not, it will derive the necessary equations and add them to the set. Essentially, this means that the only thing the user needs to provide is a starting point, namely the equation for at least one average of interest, and the order at which the equations should be truncated.

An additional noteworthy feature in this automatic completion is that the user can provide a custom filter function, which specifies whether an average should be included. This can be very useful if certain averages should be excluded from the completion algorithm since it can significantly speed up the program as symbolic rewriting operations are reduced. For example, you can use this to neglect phase-dependent terms that are actually zero in a laser model (cf. code sample 4).

### 3.4.2 Two-time correlation functions

As the name suggests, this type of correlation function depends on two different time parameters. For example, in order to compute the spectrum shown in Fig. 1(c), we need to consider the correlation function defined by

$$g(t, \tau) = \langle a^\dagger(t+\tau) a(t) \rangle. \qquad (21)$$

The spectrum is then determined by taking the Fourier transform of Eq. (21). This correlation function depends not only on the time $t$ but also on the delay $\tau$. The detailed method with which the correlation functions and spectra are computed is rather extensive, and we refer to Appendix A for details.

The basic idea is that the equation of motion for a two-time correlation function such as the one in Eq. (21) is determined by the equation of motion of the operator that depends on the delay $\tau$, which is also known as the quantum regression theorem. Deriving this equation will usually lead to couplings with other two-time correlation functions. Ultimately, one obtains a system of equations involving correlation functions, whose initial values are determined by the state of the original system under consideration at time $t$. For example, at $\tau = 0$, the correlation function in Eq. (21) is equal to the average cavity photon number at time $t$. If the state of the system is known at time $t$ one can then solve for the correlation function.

The entire procedure of finding the equations that govern a two-time correlation function and generating the code necessary to find a numerical solution is implemented in QuantumCumulants.jl with the `CorrelationFunction` type. Furthermore, spectra can be computed from correlation functions with the `Spectrum` functionality.

### 3.4.3 Mixing orders

In certain systems it might make sense to use different orders of the cumulant expansion in different context. For example, one could envision a situation where a laser model involving a cavity mode and many atoms should be treated such that the atoms are approximately independent of one another, but build up some correlations with the cavity field. Thus, it makes sense to expand all expectation values involving atomic operators only to first order, whereas cavity mode expectation values as well as cavity-atom expectation values (such as $\langle a^\dagger \sigma^{ge} \rangle$) are kept up to the second order. This approach can be quite useful since then the number of equations scales only linearly with the atom number, whereas in second order it would already scale quadratically.

Mixed orders can be used within the framework simply by providing a vector instead of a single integer as the order parameter in the cumulant expansion. Each entry in that vector corresponds to the order of the subsystem of the product space. The order to be applied to an average is then determined by the action on the respective subspace. If an average acts on multiple subsystems, the order is chosen according to a function which by default picks the maximum (note that this can be changed by the user).

# 4 Examples

In this section we will showcase the framework's usefulness by implementing a few examples. Note that similar examples can also be found in the documentation [31]. Before we dive into more involved examples, let us briefly show the one used in Sec. 2. In code sample 2, we see how one can implement a single-atom laser model in the framework.

```
using QuantumCumulants

# Define hilbert space and fundamental operators
hf = FockSpace(:cavity)
ha = NLevelSpace(:atom,(:g,:e))
h = hf ⊗ ha
@qnumbers a::Destroy(h) σ::Transition(h)

# Hamiltonian and collpase operators
@cnumbers Δ g γ κ ν
H = Δ*a'*a + g*(a'*σ(:g,:e) + a*σ(:e,:g))
J = [a,σ(:g,:e),σ(:e,:g)]
rates = [κ,γ,ν]

# Derive a set of second-order equations
eqs = meanfield(a'*a,H,J;rates=rates,order=2)
eqs_completed = complete(eqs)

# Convert to an ODESystem and solve numerically
using OrdinaryDiffEq, ModelingToolkit
@named sys = ODESystem(eqs_completed)
u0 = zeros(ComplexF64,length(eqs_completed))
p = (Δ, g, γ, κ, ν)
p0 = (0.5, 1.5, 1.25, 1, 4)
prob = ODEProblem(sys,u0,(0.0,20.0),p.=>p0)
sol = solve(prob,RK4())

# Compute the spectrum
c = CorrelationFunction(a', a, eqs_completed;
                        steady_state=true)
S = Spectrum(c,p)
ω = range(-π,π,length=301)
s = S(ω,sol.u[end],p0)
```

Code sample 2: *The single-atom laser example implemented with QuantumCumulants.jl.*

Let us stress here that the code shown in code sample 2 is actually more general than what has been discussed in Sec. 2. In particular, one could consider a higher-order approximation simply by editing one line in the script when completing the system of equations, e.g.

```
eqs_completed = complete(eqs;order=4)
```

will use a fourth-order cumulant expansion. The script will produce and solve the equations as well as compute the spectrum of the laser. In fact, the results computed in code sample 2 were used in the graphs shown in Fig. 1(b) and Fig. 1(c).

## 4.1 A laser with a three-level pump scheme

Here, we will show how one can implement systems with atoms that feature multiple energy levels. Consider therefore a slightly modified version of the single-atom laser, that features an atom with three levels. This is inspired by the first laser ever built, which used Ruby as a gain medium [37]. There are three states, which we denote by $|1\rangle, |2\rangle$ and $|3\rangle$, respectively. The atoms are incoherently driven from the ground state $|1\rangle$ to the state $|2\rangle$, which is the highest in energy, at a rate $\nu$. The state $|2\rangle$ then decays non-radiatively into the state $|3\rangle$ at a rate $\Gamma$. While the state $|3\rangle$ can also decay to the ground state $|1\rangle$ at a rate $\gamma$, population inversion can be achieved in $|3\rangle$ so long as $\Gamma, \nu \gg \gamma$. Coupling a cavity to the transition $|1\rangle$ to $|3\rangle$ allows stimulated emission and subsequent amplification thus leading to lasing action.

```
using QuantumCumulants

# Hilbert space
hf = FockSpace(:cavity)
ha = NLevelSpace(:atom, 3)
h = hf ⊗ ha

# Parameters and operators
@cnumbers Δ₃ g Γ γ κ ν
@qnumbers a::Destroy(h) σ::Transition(h)

# Hamiltonian and Decay
H = Δ₃*σ(3,3) + g*(a'*σ(1,3) + a*σ(3,1))
J = [a,σ(3,2),σ(1,3),σ(2,1)]
rates = [κ,Γ,γ,ν]

# Derive equations
eqs = meanfield([a'*a,σ(3,3),σ(2,2)],H,J;
                        rates=rates,order=4)
eqs_completed = complete(eqs)

# Solve
using OrdinaryDiffEq, ModelingToolkit
@named sys = ODESystem(eqs_completed)
u0 = zeros(ComplexF64, length(eqs_completed))
p0 = (Δ₃=>0, g=>1.8, Γ=>20, γ=>1.5, κ=>1, ν=>10)
prob = ODEProblem(sys,u0,(0.0,6.0),p0)
sol = solve(prob,RK4())
```

Code sample 3: *Three-level pump scheme in a laser.*

The Hamiltonian of this system reads

$$H = \hbar\Delta_3 \sigma^{33} + \hbar g \left(a^\dagger \sigma^{13} + a \sigma^{31}\right), \quad (22)$$

where $\Delta_3 = \omega_3 - \omega_c$ is the detuning between the cavity and the lasing transition, and $g$ is the coupling strength between the cavity and the atom. The implementation of this model is shown in code sample 3.

The key difference to the previous laser model is that we added an additional energy level to the atom by specifying the underlying Hilbertspace as `NLevelSpace(:atom, 3)`. In general, any number of energy levels can be specified here. It would thus be simple to generalize the model further to describe, for example, a gain medium with four levels.

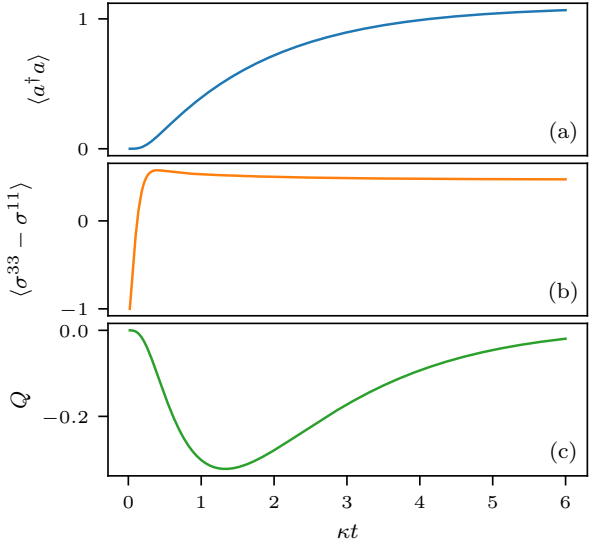

Figure 2: *Results of the three-level laser*. (a) The average photon number in the cavity and (b) the population inversion of the lasing transition and (c) the Mandel-$Q$ parameter. The parameters are $\Delta_3 = 0$, $g = 1.8\kappa$, $\Gamma = 20\kappa$, $\gamma = 1.5\kappa$ and $\nu = 10\kappa$.

We plot the average cavity photon number in Fig. 2(a) as well as the population inversion of the lasing transition in Fig. 2(b). As we can see, the system indeed forms a minimal version of a laser. In addition, we also plot the so-called Mandel-$Q$ parameter in Fig. 2(c), which is defined by

$$Q = \frac{\Delta n^2 - \langle n \rangle}{\langle n \rangle}, \qquad (23)$$

where $n = a^\dagger a$ and $\Delta n^2 = \langle n^2 \rangle - \langle n \rangle^2$ is the variance of $n$. This parameter is a measure of non-classicality in that if $Q < 0$, the photon number variance is smaller than the average photon number, which corresponds to sub-Poissonian photon statistics. Clearly, in Fig. 2(c) $Q$ is negative during the build-up phase of the laser and then tends towards 0. This signifies that a higher-order approximation (in this case moments up to fourth order are kept) suffices to capture nonclassical effects such as photon antibunching.

Let us stress here just how simple a fourth-order cumulant expansion becomes when using QuantumCumulants.jl. Deriving the 30 equations necessary for the fourth-order description would usually be challenging. With our framework, this can be achieved within a few lines of code and a runtime on the order of 10 seconds.

Note that the examples so far are relatively simple and (at least for low photon numbers) could have been treated easily using a master equation as well. The aim of the next two examples will be to examine situations where this is no longer the case.

### 4.2 Pulsed superradiant laser

```
using QuantumCumulants

# Define parameters
N = 50
@cnumbers Δ g γ κ

# Define hilbert space
hf = FockSpace(:cavity)
ha = [NLevelSpace(Symbol(:atom,i),2) for i=1:N]
h = ⊗(hf, ha...)

# Define the fundamental operators
a = Destroy(h,:a)
σ(i,j,k) = Transition(h,Symbol(:σ, k),i,j,k+1)

# Hamiltonian
H = Δ*a'*a + g*sum(a'*σ(1,2,i) + a*σ(2,1,i) for i=1:N)

# Collapse operators
J = [a;[σ(1,2,i) for i=1:N]]
rates = [κ;[γ for i=1:N]]

# Derive equations for populations
ops = [σ(2,2,i) for i=1:N]
eqs = meanfield(ops,H,J;rates=rates,order=2)

# Complete but neglect phase-dependent terms
ϕ(x::Average) = ϕ(x.arguments[1])
ϕ(::Destroy) = -1
ϕ(::Create) = 1
ϕ(x::QTerm) = sum(map(ϕ, x.args_nc))
ϕ(x::Transition) = x.i - x.j
phase_invariant(x) = iszero(ϕ(x))
complete!(eqs;filter_func=phase_invariant)

using OrdinaryDiffEq, ModelingToolkit
@named sys = ODESystem(eqs)
u0 = zeros(ComplexF64, length(eqs))
u0[1:N] .= 1.0 # atoms are inverted initially
p0 = (Δ=>0.5, g=>0.5, γ=>0.25, κ=>1)
prob = ODEProblem(sys,u0,(0.0,10.0),p0)
sol = solve(prob,RK4())
```

Code sample 4: *Laser model with multiple atoms as gain medium.*

The previous examples showed comparably simple models of a laser where the gain medium consisted of only one atom. Here, we will consider a cavity mode which couples to a number of atoms $N$, which are initially inverted such that they provide gain. Note that we label the atomic ground state by $|1\rangle$ and the excited state by $|2\rangle$. The coherent dynamics is described by the Tavis-Cummings Hamiltonian,

$$H = \hbar \Delta a^\dagger a + \sum_{j=1}^{N} \hbar g_j \left( a^\dagger \sigma_j^{12} + a \sigma_j^{21} \right), \qquad (24)$$

where $g_j$ is the coupling rate of the $j$th atom. In addition, all atoms are subject to spontaneous emission at the rate $\gamma$. The implementation of this system in a second-order approximation for $N = 50$ is shown in code sample 4.

Note, that in the implementation we assumed equal couplings, i.e. $g \equiv g_j$. This is simply due to keeping the displayed code short and one can easily generalize code sample 4 to individual couplings. As can be seen in Fig. 3(a), a gain medium featuring $N = 50$ atoms leads to a substantial superradiant pulse. After an initial brief build-up of atomic coherences, collective stimulated emission leads to a steep increase of photons in the cavity mode. Following the initial pulse we observe oscillations in which the atoms reabsorb photons from the cavity [see Fig. 3(b)] and emit them again. The pulse size reduces over time as photons leak out of the cavity until finally no photons and atomic excitations are left. The results shown in Fig. 3 are in good qualitative agreement with the experimental demonstration of such a pulsed superradiant laser in Ref. [21].

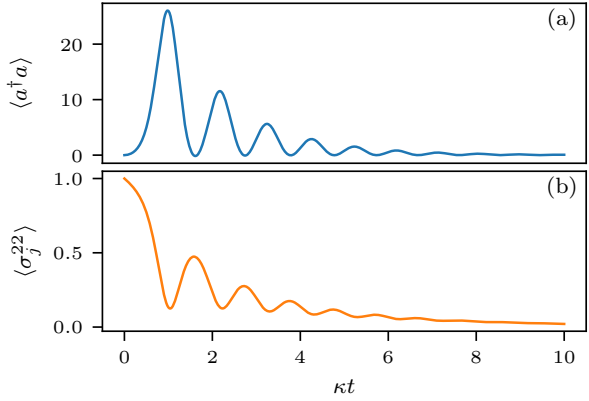

Figure 3: *Pulsed superradiant laser*. (a) The average photon number and (b) the excited state population of an atom inside the gain medium for a cavity mode coupled to $N = 50$ atoms that are initially in the excited state. The parameters are $\Delta = g = \kappa/2$, $\gamma = \kappa/4$ and $\nu = 4\kappa$.

Note that the key point of this example is that we are treating a number of atoms that could not be investigated easily in a master equation approach as the size of the Hilbert space would already be huge. Deriving the necessary equations for a second-order approximation with our framework and a subsequent numerical solution takes only minutes on any modern computer and does not require a significant amount of memory. Increasing the atom number further to, say $N = 100$, code sample 4 would still finish in a time on the order of minutes, whereas a treatment with a master equation would be downright impossible at that point. The limiting factor for the cumulant expansion approach ultimately is the number of equations. Specifically, the number of equations in code sample 4 scales with the number of atoms as

$$n_\text{eqs} = \frac{N(N-1)}{2} + 2N + 1. \quad (25)$$

For $N = 50$, as used in code sample 4, this results in $n_\text{eqs} = 1326$ equations that need to be derived and solved.

As mentioned above, one could generalize the example to feature individual couplings of the atoms to the cavity or dissipation rates of the atomic ensemble. While we assumed ideal atomic positioning here, one could therefore also include fluctuations in the positions and hence the coupling rates. To do so, you would need to compute the time evolutions for many randomly generated atomic configurations and average over the results. So long as the time evolution itself remains deterministic, our toolbox is perfectly suitable for such simulations. It is important to note that in such a case the computationally intense symbolic derivation of equations still only has to be done once. Numerical variations of parameters can be done when setting the values for the time evolution.

### 4.3 Optomechanical cooling of a micromechanical oscillator

In this example, we show how to implement a cooling scheme based on radiation pressure coupling of light to a mechanical oscillator, such as a membrane. The oscillator is placed inside an optical cavity. The cavity is driven by a laser and the resulting radiation pressure of the cavity field effectively couples the photons in the cavity mode to the vibrational phonons of the mechanical oscillator mode. This model is based on the one studied in Ref. [38], and the Hamiltonian reads

$$H = -\hbar\Delta a^\dagger a + \hbar\omega_\text{m} b^\dagger b + \hbar G a^\dagger a \left(b + b^\dagger\right) + \\ \hbar E \left(a + a^\dagger\right), \quad (26)$$

where $\Delta = \omega_\ell - \omega_c$ is the detuning between the laser ($\omega_\ell$) driving the cavity ($\omega_c$). The amplitude of the laser is denoted by $E$, the resonance frequency of the mechanical oscillator by $\omega_m$, and the radiation pressure coupling is given by $G$. The operators $a$ and $b$ are the photonic and phononic annihilation operators, respectively. Additionally, photons leak out of the cavity at a rate $\kappa$.

We will consider the membrane at room temperature. Its vibrational mode is in a thermal state with an average number of phonons that can be estimated from

$$k_B T = n_\text{vib} \hbar \omega_m. \quad (27)$$

If the resonator has a resonance frequency of $\omega_m = 10\text{MHz}$, then the number of phonons at room temperature ($T \approx 300K$) is approximately $n_\text{vib} \approx 4 \times 10^6$. Let us stress here that treating such a large number of phonons in a master equation approach is problematic since this number determines the cut-off which can be chosen and thus the dimension of the Hilbert space. The averages treated in the cumulant expansion approach, however, are independent of this cut-off. Therefore, arbitrarily large numbers can be used (of course numerical floating-point errors may become substantial at some point). This makes QuantumCumulants.jl an ideal candidate to treat optomechanical problems in a high-temperature regime.

```julia
using QuantumCumulants
using OrdinaryDiffEq, ModelingToolkit

# Hilbert space
hc = FockSpace(:cavity)
hm = FockSpace(:motion)
h = hc ⊗ hm

# Operators
@qnumbers a::Destroy(h,1) b::Destroy(h,2)

# Parameters
@cnumbers Δ ωm E G κ

# Hamiltonian
H = -Δ*a'*a + ωm*b'*b + G*a'*a*(b + b') + E*(a + a')

# Derive equations
eqs = meanfield([b'*b,a'*a],H,[a];rates=[κ],order=2)
eqs_completed = complete(eqs)

# Numerical solution
@named sys = ODESystem(eqs_completed)
u0 = zeros(ComplexF64, length(eqs_completed))
u0[1] = 4*1e6 # Initial number of phonons
p0 = (Δ=>-10, ωm=>1, E=>200, G=>0.0125, κ=>20)
prob = ODEProblem(sys,u0,(0.0,60000),p0)
sol = solve(prob,RK4())
```

Code sample 5: *Cooling of a micromechanical oscillator at room temperature.*

The code implementing this model in a second-order approximation is shown in code sample 5. Note, that in order to accurately represent a thermal state, we cannot treat the problem in a first-order approximation since, in a thermal state, $\langle b \rangle = 0$, but the number of phonons is $\langle b^\dagger b \rangle \neq 0$ (with the initial value given by $n_\text{vib}$).

In Fig. 4 we show how the temperature of the membrane reduces while the photon number inside the cavity builds up. The cooling works quite well, such that a final temperature below 1mK is reached.

## 5 Limitations and Outlook

As we have shown by now, the framework can be quite useful. However, there are some disadvantages, which we want to discuss in this section.

### 5.1 Principal limitations

First, we will provide an overview of the fundamental limits inherent to the approach. Mean-field descriptions based on the cumulant expansion are approximations that neglect a certain amount of quantum correlations. As such, they are not well suited to treat problems featuring large degrees of entanglement. In principle, one could go to ever higher orders to overcome this. However, this will polynomially increase the number of equations and quickly becomes unfeasible. Furthermore, there are cases where a full quantum treatment is simply necessary, and thus the mean-field approach cannot be used.

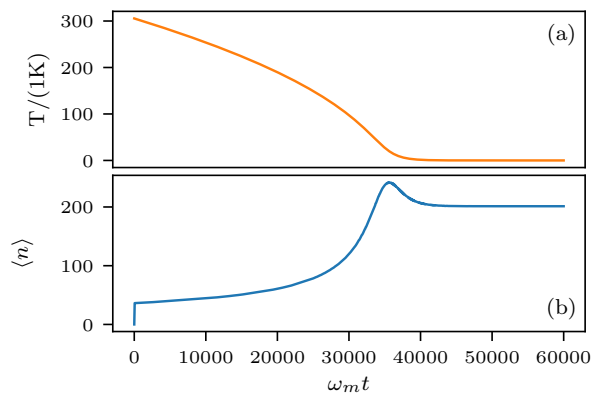

Figure 4: *Optomechanical cooling of a membrane.* (a) The temperature of the membrane reduces from room temperature over time. (b) At the same time, the field in the cavity builds up accumulating photons. The parameters were $\omega_m = 10\text{MHz}$, $\Delta = -10\omega_m$, $E = 200\omega_m$, $G = \omega_m/80$ and $\kappa = 20\omega_m$.

This brings us to a related issue: there is no generic way to know for certain if neglecting joint cumulants above a specific order is sufficiently accurate. The assumption that neglected cumulants are zero in principle has to be checked for each system and parameter set considered. This can only be done by actually computing the necessary cumulants or checking for convergence when going to higher orders. Note that this is not always possible as a higher order means having to solve more equations.

When considering large composite systems, the number of equations increases polynomially, where the leading order of the polynomial is determined by the order. For example, $N$ two-level atoms in second order lead to $\mathcal{O}(N^2)$ equations, in third order to $\mathcal{O}(N^3)$, etc. While this is a huge improvement over the exponential scaling with $2^N$ in a full quantum treatment, it can still limit the treatment of many atoms in higher orders.

### 5.2 Current limitations

There are some disadvantages to the framework that can possibly be addressed by changes in the future. One example is that QuantumCumulants.jl is currently not able to handle expressions involving exponentials of operators or similar functions. The exponential of an operator represents an infinite series involving all powers of its argument. As such, it is not straightforward to form the cumulant expansion. While it can be done for specific expressions, attempts at finding a generic way to do this for products involving multiple exponentials and other operators have not been successful so far.

Two more restrictions of QuantumCumulants.jl are not inherent, but the need to address them did not arise so far. The fact that we choose normal ordering of expressions is somewhat arbitrary and could be changed if, for example, one has a use-case that requires anti-normal ordering. Finally, there are currently only two Hilbert spaces implemented, namely `FockSpace`s and `NLevelSpace`s. Adding in different Hilbert spaces can be done, of course. We note, though, that a surprisingly large number of problems can be treated with only the two Hilbert spaces currently implemented.

### 5.3 Future changes

While QuantumCumulants.jl has already proven very useful, it is still in an early stage of development. Things may still change and new features will be added. Here we want to comment on the latter and mention a few possible directions.

At the very beginning when reviewing the theory, we presented the Quantum Langevin equation (QLE), Eq. (2), which features quantum noise. Throughout the entire paper, this noise was neglected as it does not contribute to averages. One possible future development is to actually consider this noise explicitly.

On the one hand, this would open the possibility to implement an approach based on the linearization of the operator equations around large average values. In steady state one can then compute the covariance matrix from a first-order (mean-field) solution of the system. This approach was used, for example, in Ref. [38].

On the other hand, one can convert the QLEs to $c$-number Langevin equations [39, 40], i.e. a set of stochastic differential equations. When the noise processes are chosen such that they reproduce the actual correlation functions of the quantum noise, one can obtain the second-order solutions by averaging over many trajectories. Additionally, one does not need to rely on white noise assumptions, but colored noise could be considered as well. Note that the ModelingToolkit.jl [34] framework and the DifferentialEquations.jl [35, 41] library are already capable of handling stochastic differential equations.

Both the approaches based on the treatment of noise allow the reconstruction of second-order expectation values from a first-order solution. As such, they can be used in systems which are too large to be treated even in second order, but still exhibit some quantum phenomena.

Finally, another useful method exploits symmetry properties of multiple subsystem in order to significantly reduce the number of equations. This is done, for example, in the lasing model in Ref. [17] where atoms in the gain medium are assumed to be identical, which allows for an effective description with only a few equations for an arbitrarily large gain medium. An experimental automized version of this approach is already implemented in QuantumCumulants.jl (see Appendix B).

## 6 Conclusions

We developed a novel open framework called QuantumCumulants.jl written in Julia, which automizes the analytical calculations and numerical implementation required for generalized mean-field approximations of the dynamics of open quantum systems. To this end we combined the symbolic handling of noncommutative variables for the automated cumulant expansion of moments to a specified order with other libraries offering automated code generation suitable for direct numerical solutions by standard differential equation solvers. After showing the general principle at the hand of the generic single atom laser model, we demonstrated the usefulness and versatility of our toolbox in several examples.

The framework already shows promising capabilities. Yet, the implications that it works at all go way beyond what can currently be done: the cumulant expansion approach is a powerful tool in its own right. Making the steps behind it easily accessible will allow a swift treatment of numerous problems that cannot be treated in a full quantum model. As development of QuantumCumulants.jl continues, the number of problems that can be addressed will continue to increase. Furthermore, even at points where the number of equations becomes too large to be handled by hand, the toolbox can still be successful. Potentially, the framework could be applied to problems not only in quantum optics, but also in quantum information, nanophotonics, and even condensed matter theory.

### Acknowledgments

We would like to thank Anna Bychek and Georgy Kazakov for helpful discussions and testing of the software. We acknowledge funding from the European Union's Horizon 2020 research and innovation program under Grant Agreement No. 820404 iqClock. The graphs were produced with the open-source library Matplotlib [42].

# A Correlation functions and power spectra

## A.1 Correlation functions

Correlation functions can provide useful information about a system. The expectation values used to compute a correlation function are special in that they generally depend on two different times. Consider two operators $A$ and $B$. Their two-time correlation function is given by

$$C(t,\tau) = \langle A(t+\tau)B(t) \rangle. \qquad (28)$$

If the time difference $\tau = 0$, the correlation function is just the expectation value $\langle AB \rangle$ at time $t$. So in order to compute the correlation function, we can evolve a system of equations containing the expectation value at $\tau = 0$ up to a time $t$. The evolution with respect to the time delay is then determined by

$$\frac{d}{d\tau}C(t,\tau) = \langle \left(\frac{d}{d\tau}A(t+\tau)\right)B(t)\rangle. \qquad (29)$$

Hence, we can derive the set of equations required to compute the correlation function from the equation of the operator $A$. Then, the same procedure as for a standard time evolution is used: the set of equations is expanded to a certain order and completed. Numerical code to solve the underlying differential equations is automatically generated.

As there is quite a bit of new theory involved here, let us go back to the example used throughout Sec. 2, namely the single-atom laser, to clarify the procedure explained above. The first-order coherence function of the cavity field is given by

$$C(t,\tau) = \langle a^\dagger(t+\tau)a(t) \rangle. \qquad (30)$$

The equations of motion with respect to $\tau$ are readily constructed [cf. Eqs. (4)],

$$\frac{d}{d\tau}\langle a^\dagger(t+\tau)a(t)\rangle = \left(i\Delta - \frac{\kappa}{2}\right)\langle a^\dagger(t+\tau)a(t)\rangle$$
$$+ ig\langle \sigma^{eg}(t+\tau)a(t)\rangle \qquad (31a)$$
$$\frac{d}{d\tau}\langle \sigma^{eg}(t+\tau)a(t)\rangle = -\frac{\gamma+\nu}{2}\langle \sigma^{eg}(t+\tau)a(t)\rangle$$
$$- ig\langle a^\dagger(t+\tau)\sigma^{ee}(t+\tau)a(t)\rangle. \qquad (31b)$$

Using the cumulant expansion to second order together with the phase invariance of the system, we obtain for the last term

$$\langle a^\dagger(t+\tau)\sigma^{ee}(t+\tau)a(t)\rangle = \langle a^\dagger(t+\tau)a(t)\rangle\langle \sigma^{ee}(t+\tau)\rangle. \qquad (32)$$

In principle, we would have to compute the time evolution for $\langle \sigma^{ee}(t+\tau)\rangle$ with respect to $\tau$ together with the other equations. However, we will consider the system to be in steady state in order to avoid having to treat this additional equation.

## A.2 Steady state

If the original system is evolved up to a time $t$ such that it is in steady state, i.e. expectation values no longer change after that time, the set of equations determining the correlation function has a special property. Specifically, after the cumulant expansion has been performed, there can only be a single term in each product on the right-hand-side of the set of equations that depends on $\tau$. All other terms depend on $t$ alone, meaning that they are constant since they no longer change after the time $t$. Therefore, the system of equations from which the correlation function is computed is linear, in the sense that it can be written as

$$\frac{d}{d\tau}\mathbf{y}(\tau) = \mathbf{M}\mathbf{y}(\tau) + \mathbf{d}, \qquad (33)$$

where $\mathbf{y}(\tau)$ is the vector of $\tau$-dependent variables. The elements of the matrix $\mathbf{M}$ as well as the vector $\mathbf{d}$ are given by steady-state expectation values and parameters, i.e. they are independent of $\tau$.

In the case of the single-atom laser example, we have $\mathbf{y}(\tau) = \left(\langle a^\dagger(t+\tau)a(t)\rangle, \langle \sigma^{eg}(t+\tau)a(t)\rangle\right)^T$, $\mathbf{d} = \mathbf{0}$, and

$$\mathbf{M} = \begin{pmatrix} i\Delta - \frac{\kappa}{2} & ig \\ -ig\langle \sigma^{ee}\rangle & -\frac{\gamma+\nu}{2} \end{pmatrix}. \qquad (34)$$

## A.3 Power spectra

According to the Wiener-Khinchin theorem. the spectral density associated with a correlation function is given by its Fourier transform,

$$S(t,\omega) = 2\mathrm{Re}\left\{\int d\tau e^{-i\omega\tau}C(t,\tau)\right\}. \qquad (35)$$

In order to compute this, we can solve the system of equations determining $C(t,\tau)$, subsequently taking the Fourier transform. However, if we are not interested in the temporal behavior of the correlation function, and if the system of which we want to compute the spectrum is in steady state, we can directly compute the spectrum from Eq. (33). To this end, we define

$$\mathbf{x}(s) = \mathcal{L}\{\mathbf{y}(\tau)\}, \qquad (36)$$

where $\mathcal{L}$ denotes the Laplace transform with respect to $\tau$. Taking the Laplace transform of Eq. (33), we have

$$(s\mathbb{1} - \mathbf{M})\mathbf{x}(s) = \mathbf{y}(0) + \frac{\mathbf{d}}{s}. \qquad (37)$$

Note that the Laplace transform is equivalent to the Fourier transform at the point where $s = i\omega$, i.e $S(\omega) = 2\mathrm{Re}\{\mathbf{x}_1(i\omega)\}$. Hence, instead of computing the time evolution of the correlation function we can directly compute the spectrum by solving the linear equation

$$\mathbf{x} = \mathbf{A}^{-1}\mathbf{b}, \qquad (38)$$

where $\mathbf{A} = i\omega\mathbb{1} - \mathbf{M}$ and $\mathbf{b} = \mathbf{y}(0) + \mathbf{d}/(i\omega)$.

For the single-atom laser, solving Eq. (38) requires computing the inverse of a simple $2 \times 2$ matrix. For larger systems, the method using a Laplace transform is usually faster than integrating a system of equations of the same size. Additionally, it avoids numerical errors of the integration and the subsequent discrete Fourier transform.

## B  Dealing with many identical subsystems

Consider a system consisting of multiple subsystems. If a subset of these subsystems is guaranteed to be identical at any point in time the number of equations needed to describe the whole system can be significantly reduced. In essence, instead of explicitly describing each subsystem, we may only describe a single one and place appropriate scaling factors at some points in the equations of motion.

For clarity, take the example from Sec. 4.2. The equation of motion for the average field amplitude is

$$\langle \dot{a} \rangle = -\left(i\Delta + \frac{\kappa}{2}\right)\langle a \rangle + \sum_{j=1}^{N} g_j \langle \sigma_j^{12} \rangle. \quad (39)$$

Now, given that all atoms are initially in the same state and their couplings to the cavity mode and the environment are all equal (i.e. $g_j \equiv g \; \forall \; j$ and similarly for $\gamma_j$ and $\nu_j$), then we know that $\langle \sigma_j^{k\ell} \rangle \equiv \langle \sigma_1^{k\ell} \rangle \; \forall \; j$. We may then rewrite the above equation and obtain

$$\langle \dot{a} \rangle = -\left(i\Delta + \frac{\kappa}{2}\right)\langle a \rangle + Ng \langle \sigma_1^{12} \rangle. \quad (40)$$

Thus, we only need to know $\langle \sigma_1^{12} \rangle$ in order to solve the above equation, instead of all the different $\langle \sigma_j^{12} \rangle$. Things are a bit more tricky, however, when looking at the equations of motion for atomic expectation values or mixed atom-field expectation values. One has to take care such that the equations reproduce the correct correlations between individual atoms. For example, the complete set of second-order equations of the laser model involving many atoms reads

$$\frac{d}{dt}\langle \sigma_1^{22} \rangle = ig\langle a^\dagger \sigma_1^{12} \rangle - ig\langle a\sigma_1^{21} \rangle - \gamma\langle \sigma_1^{22} \rangle \quad (41a)$$

$$\frac{d}{dt}\langle a^\dagger \sigma_1^{12} \rangle = ig\langle \sigma_1^{22} \rangle - ig\langle a^\dagger a \rangle + i\Delta\langle a^\dagger \sigma_1^{12} \rangle - 0.5\gamma\langle a^\dagger \sigma_1^{12} \rangle - 0.5\kappa\langle a^\dagger \sigma_1^{12} \rangle + 2ig\langle \sigma_1^{22} \rangle\langle a^\dagger a \rangle \quad (41b)$$
$$+ ig\langle \sigma_1^{21}\sigma_2^{12} \rangle(-1 + N)$$

$$\frac{d}{dt}\langle a^\dagger a \rangle = -\kappa\langle a^\dagger a \rangle - iNg\langle a^\dagger \sigma_1^{12} \rangle + iNg\langle a\sigma_1^{21} \rangle \quad (41c)$$

$$\frac{d}{dt}\langle \sigma_1^{21}\sigma_2^{12} \rangle = ig\langle a^\dagger \sigma_1^{12} \rangle - ig\langle a\sigma_1^{21} \rangle - \gamma\langle \sigma_1^{21}\sigma_2^{12} \rangle - 2ig\langle \sigma_1^{22} \rangle\langle a^\dagger \sigma_1^{12} \rangle + 2ig\langle \sigma_1^{22} \rangle\langle a\sigma_1^{21} \rangle \quad (41d)$$

As we can see, to fully capture the correlations we need to consider not a single atom, but two. Another noteworthy thing is that the mixed atom-field expectation value Eq. (41b) couples to all but one of the atomic ensemble. Therefore, the correlations are multiplied by $N - 1$.

This approach is very useful to study atom numbers that would be otherwise inaccessible in even a first- or second-order description. The key point is that the number of equations is independent of the number of identical subsystems. The set of equations shown above is similar to the one studied in Ref. [17], where the symmetry property of the gain medium of a superradiant laser has been exploited to the same end as shown here.

QuantumCumulants.jl implements an experimental version that automates the procedure of reducing equations and finding the correct scaling factors. To this end, we need to find the positions of the sums that occur when dealing with each subsystem explicitly. On top of that, sums may exclude certain indices which changes the scaling factor. Ultimately, we devised a set of rules that obtains this information based on the action of operators on the different Hilbert spaces. Similar rules are implemented to judge whether an expectation value can be eliminated by substitution. For example, whenever $\langle \sigma_2^{ij} \rangle$ occurs in the above equations, we can replace it by $\langle \sigma_1^{ij} \rangle$.

Currently, this functionality is tailored to Hamiltonians of the form similar to Eq. (24). To specify, we can currently treat systems comprised of many identical subsystems that are coupled by only one common

Hilbert space. In the example shown here, the identical subsystems are the atoms inside the gain medium and they couple via the common cavity mode. Note that the approach is not restricted to this combination of Hilbert spaces, i.e. we could also treat a single atom coupling to many identical modes, or a single designated mode coupling to other modes. However, this automatic procedure is not applicable to more general problems and will be subject to changes in the future. Hence, we consider this part of the framework experimental.

The program which derives Eqs. (41) automatically and solves them is shown in code sample 6. This requires the use of a `ClusterSpace` which represents an arbitrary amount of identical copies of a single Hilbert space. Note that we also need to make the order of the problem known to the space in the beginning, as this is the actual number of Hilbert spaces required to describe the system properly. Another advantage of this approach is that the equations only need to be derived once after which one can solve the system dynamics for arbitrary numbers of atoms as $N$ is simply a parameter in code sample 6.

```julia
using QuantumCumulants

M = 2 #order
@cnumbers N Δ g κ γ
hf = FockSpace(:cavity)
ha1 = NLevelSpace(:atom, 2)
ha = ClusterSpace(ha1, N, M)
h = tensor(hf, ha)

@qnumbers a::Destroy(h)
# (i,j) -> levels, k -> atomic index
σ(i,j,k) = Transition(h, :σ, i, j)[k]

# Hamiltonian
H = Δ*a'*a + g*sum(a'*σ(1,2,i) + a*σ(2,1,i) for i=1:M)

# Collapse operators
J = [a;[σ(1,2,i) for i=1:M]]
rates = [κ;[γ for i=1:M]]

# Derive equation for atomic population
eqs = meanfield(σ(2,2,1),H,J;rates=rates,order=M)

# Complete but neglect phase-dependent terms
ϕ(x::Average) = ϕ(x.arguments[1])
ϕ(::Destroy) = -1
ϕ(::Create) = 1
ϕ(x::QTerm) = sum(map(ϕ, x.args_nc))
ϕ(x::Transition) = x.i - x.j
phase_invariant(x) = iszero(ϕ(x))
complete!(eqs;filter_func=phase_invariant)

using OrdinaryDiffEq, ModelingToolkit
@named sys = ODESystem(eqs)
u0 = zeros(ComplexF64, length(eqs))
u0[1] = 1.0 # atoms are inverted intially
p0 = (Δ=>0.5, g=>0.01, γ=>0.25, κ=>1, N=>500000)
prob = ODEProblem(sys,u0,(0.0,20.0),p0)
sol = solve(prob,RK4())
```

Code sample 6: *Automatic elimination of identical subsystems.* The above code derives and solves the equations for $N = 5 \times 10^5$ atoms in the gain medium.